\documentclass[useAMS,usenatbib]{mn2e}
\usepackage{graphicx}
\usepackage{color}
\usepackage{soul} 
\usepackage{url}
\usepackage{txfonts}
\usepackage{longtable}
\usepackage[normalem]{ulem}

\title[Chemical abundances of symbiotic giants -- II]{
Chemical abundance analysis of symbiotic giants -- II. AE\,Ara, BX\,Mon, KX\,TrA, and CL\,Sco}
\author[C. Ga{\l}an et al.]{Cezary Ga{\l}an$^{1}$\thanks{E-mail: cgalan@camk.edu.pl}, Joanna Miko{\l}ajewska$^{1}$, and Kenneth H. Hinkle$^{2}$ 
\\
$^{1}$N. Copernicus Astronomical Center, Bartycka 18, PL-00-716 Warsaw, Poland\\
$^{2}$National Optical Astronomy Observatory, PO Box 26732, Tucson, AZ 85726, USA}

\begin{document}

\date{Accepted 2014 November 12. Received 2014 October 10}

\pagerange{\pageref{firstpage}--\pageref{lastpage}} \pubyear{2014}

\maketitle

\label{firstpage}

\begin{abstract}
Knowledge of the elemental abundances of symbiotic giants is essential to
address the role of chemical composition in the evolution of symbiotic
binaries, to map their parent population, and to trace their mass transfer
history.  However, there are few symbiotic giants for which the photospheric
abundances are fairly well determined.
This is the second in a series of papers on chemical composition of
symbiotic giants determined from high-resolution (R\,$\sim$\,50000) near-IR
spectra.  Results are presented for the late-type giant star in the AE\,Ara,
BX\,Mon, KX\,TrA, and CL\,Sco systems.  Spectrum synthesis employing
standard local thermal equilibrium (LTE) analysis and stellar atmosphere
models were used to obtain photospheric abundances of CNO and elements
around the iron peak (Sc, Ti, Fe, and Ni).
Our analysis resulted in sub-solar metallicities in BX\,Mon, KX\,TrA, and
CL\,Sco by [Fe$/$H]$\sim-0.3$ or $-0.5$ depending on the value of
microturbulence.  AE\,Ara shows metallicity closer to solar by
$\sim0.2$\,dex.  The enrichment in $^{14}$N isotope found in all these
objects indicates that the giants have experienced the first dredge-up.  In
the case of BX\,Mon first dredge-up is also confirmed by the low
$^{12}$C/$^{13}$C isotopic ratio of $\sim$8.

\end{abstract}


\section{Introduction}

Symbiotic stars are long-period binary systems composed of two evolved and
strongly interacting stars: a red giant donor and a hot luminous white dwarf
companion (occasionally replaced by a neutron star) surrounded by an ionized
nebula.  Mass exchange between the binary system members is critical in
defining their evolution.  Mass-loss from the giant undergoes accretion to
the compact object via wind and$/$or Roche lobe overflow (\citealt{Pod2007},
\citealt{Mik2012}) resulting in the formation of an accretion disc and jet
(\citealt{Sol1985}, \citealt{Tom2003}, \citealt{Ang2011}).  The hot
companion had previously passed through a red giant stage.  In the previous
red giant stage mass was transferred from this star to the star that is
currently a red giant.  Abundance signatures tracing this mass transfer
process have been measured in some red giant--white dwarf binary systems
\citep{SmLa1988}.  In some cases the mass transfer process can result in
symbiotic progenitors for supernovea Type Ia (SNe\,Ia).  Symbiotic systems
are believed responsible for between a few per cent to 30 per cent of
SNe\,Ia events (\citealt{Dil2012}, \citealt{Mik2012}).

\begin{table*}
 \centering
  \caption{Journal of spectroscopic observations. Velocity parameters$\,^a$
of the cool components obtained via cross-correlation technique and orbital
phases calculated according to the literature ephemeris are also
shown.}\label{T1}
  \begin{tabular}{@{}lcccccc@{}}
  \hline
           & Sp.\,Region             & Date       & HJD(mid)     & $(V_{\rmn{rot}}^2 \sin^2{i} + \xi^2_{\rmn{t}})^{0.5}$ & $V_{\rmn{rad}}$ & Orbital phase$^{b}$\\
           & Band($\lambda$[$\mu$m]) &        &           &\\
 \hline
           & $H$($\sim$1.56)         & 17.02.2003 & 2452687.8830 & ~7.76 & ~-6.34 $\pm$ 0.56  & 0.05 \\
 AE\,Ara   & $K$($\sim$2.23)         & 20.04.2003 & 2452749.8669 & ~9.16 & ~-4.47 $\pm$ 0.54  & 0.13 \\
           & $K$($\sim$2.23)         & 03.04.2004 & 2453098.8487 & 10.35 & -15.30 $\pm$ 0.67  & 0.56 \\
 \hline
           & $H$($\sim$1.56)         & 16.02.2003 & 2452686.7409 & ~6.08 & ~25.36 $\pm$ 0.48  & 0.30 \\
 BX\,Mon   & $K$($\sim$2.23)         & 20.04.2003 & 2452749.5231 & ~7.58 & ~24.12 $\pm$ 0.40  & 0.35 \\
           & $K_{\rm r}$($\sim$2.36) & 03.04.2006 & 2453828.5095 & ~8.44 & ~27.33 $\pm$ 0.56  & 0.20 \\
 \hline
           & $H$($\sim$1.56)         & 17.02.2003 & 2452687.8230 & ~6.05 & -125.09 $\pm$ 1.92 & 0.73 \\
 KX\,TrA   & $K$($\sim$2.23)         & 20.04.2003 & 2452749.7670 & ~6.29 & -126.13 $\pm$ 0.87 & 0.78 \\
           & $K$($\sim$2.23)         & 03.04.2004 & 2453098.7314 & ~5.74 & -122.20 $\pm$ 1.29 & 0.03 \\
 \hline
           & $H$($\sim$1.56)         & 17.02.2003 & 2452687.8341 & ~7.02 & -31.90 $\pm$ 0.68  & 0.07 \\
 CL\,Sco   & $K$($\sim$2.23)         & 20.04.2003 & 2452749.7780 & ~6.99 & -32.99 $\pm$ 0.36  & 0.17 \\
           & $K$($\sim$2.23)         & 15.08.2003 & 2452866.5367 & ~8.52 & -32.94 $\pm$ 0.50  & 0.36 \\
           & $K$($\sim$2.23)         & 03.04.2004 & 2453098.7794 & ~8.83 & -46.51 $\pm$ 0.80  & 0.73 \\
\hline
\end{tabular}   
\begin{list}{}{}
\item[$^{a}$]\,Units $\rmn{km}\,\rmn{s}^{-1}$.
\item[$^{b}$]\,Orbital phase calculated from the following ephemerides: AE\,Ara 2453449+803.4$\times$E \citep{Fek2010},
BX\,Mon 2449796+1259$\times$E \citep{Fek2000}, KX\,TrA 2453053+1350$\times$E
\citep{Fer2003}, CL\,Sco 2452018+625$\times$E \citep{Fek2007}. The zero-point corresponds to the inferior conjunction of the red giant.
\end{list} 
\end{table*}

The complex structure with many types of interactions make symbiotic stars
excellent laboratories for studying various aspects of red giant branch
(RGB)/ asymptotic giant branch (AGB) binary evolution.  Knowledge of the
chemical composition in symbiotic giant's atmospheres can be used to track
the mass exchange history as well as the population origin of the stellar
material.  However, reliable determinations of photospheric compositions
exist for only a small number of objects, mostly G- or K-type giants,
whereas the vast majority of symbiotic stars contain M-type giants.  Prior
to the current series of papers only four M giants in S-type symbiotic
systems had been analysed in the literature: V2116\,Oph \citep{Hin2006},
T\,CrB, RS\,Oph \citep{Wal2008}, and CH\,Cyg \citep{Sch2006}.  All of them
had solar or nearly solar metallicities.  The rarer symbiotic stars
containing K-type giants are metal-poor with s-process elements overabundant
(\citealt[1997]{Smi1996}; \citealt{Per1998}; \citealt{Per2009}) whereas
those with G-type giants have solar metallicity and s-process enhancement
(\citealt{Smi2001a}; \citealt{Per2005}).

This is the second in a series of papers on the chemical abundance analysis
of the symbiotic giants.  \citet[][hereafter Paper\,I]{Mik2014} discuss
additional motivations for this work as well as the abundance analysis for
the M star in two classical S-type symbiotic systems, RW\,Hya and SY\,Mus. 
In this paper, we obtain photospheric abundances for the M giant in four
more symbiotic systems: AE\,Ara, BX\,Mon, KX\,TrA, and CL\,Sco.  We also
make a concise comparative analysis of our present and previous results.

\begin{table}
 \centering
\caption{Stellar parameters, $T_{\rm{eff}}$ and $\log{g}$ estimated from
spectral types and $T_{\rm{eff}}$--$\log{g}$--colour relation.}
  \label{T2}
  \begin{tabular}{@{}lcccc@{}}
  \hline
                                     & AE\,Ara         & BX\,Mon         & KX\,TrA         & CL\,Sco         \\
  \hline
Sp. Type$^{[1]}$                     & M5.5            & M5              & M6              & M5              \\
$T_{\rm{eff}}$[K]$^{[2]}$            & 3300\,$\pm$\,75 & 3355\,$\pm$\,75 & 3240\,$\pm$\,75 & 3355\,$\pm$\,75 \\
$T_{\rm{eff}}$[K]$^{[3]}$            & 3312            & 3367            & 3258            & 3367            \\
$J-K$$^{[4,5]}$                      & 1.36            & 1.37            & 1.39            & 1.29            \\
$E$($B-V$)$^{[6]}$                   & 0.19--0.25      & 0.12--0.16      & 0.13--0.23      & 0.26--0.34      \\
($J-K$)$_0$                          & $\sim$1.25      & $\sim$1.29      & $\sim$1.3       & $\sim$1.15      \\
$T_{\rm{eff}}$[K]$^{[7]}$            & $<$3500         & $<$3500         & $<$3500         & 3500--3600      \\
$\log{g}$$^{[7]}$                    & $<$0.39         & $<$0.39         & $<$0.39         & $\sim$0.49      \\
$T_{\rm{eff}}$[K]$^{a}$              & 3300            & 3400            & 3300            & 3400            \\
$\log{g}$$^{a}$                      & 0.0             & 0.0             & 0.0             & 0.5             \\
$\rm {M_{\rm{bol}}}$$^{b}$           & -3.6            & -4.1            & -3.9            & $\sim$-3.0      \\
  \hline
\end{tabular}
\begin{list}{}{}
\item[{References:}] spectral types are from $^{[1]}$\cite{Mue1999},
total Galactic extinction adopted according to $^{[6]}$\cite{Sch2011} and
\cite{Sch1998}, infrared colours are from 2MASS $^{[4]}$\citep{Phi2007}
transformed to $^{[5]}$\cite{BeBr1988} photometric system.
\item[{Callibration by:}] $^{[2]}$\cite{Ric1999},
$^{[3]}$\cite{VBe1999}, $^{[7]}$\cite{Kuc2005}.
\item[{$^{a}$:}]\,adopted.
\item[{$^{b}$:}]\,based on known radii and/or pulsation properties.
\end{list}
\end{table}

\section[]{Observations and data reduction}

High-resolution ($R = \lambda/\Delta\lambda \sim 50000$), high-S/N ratio
($\ga$\,100), near-IR spectra were observed with the Phoenix cryogenic
echelle spectrometer on the 8\,m Gemini South telescope in 2003--2006.  All
the spectra cover narrow spectral intervals ($\sim$100\AA) located in the
$H$ and $K$ photometric bands at mean wavelengths 1.563, 2.225, and
2.361\,$\mu$m (hereafter $H$, $K$, and $K_{\rm r}$-band spectra,
respectively).  The $H$-band spectra contain molecular CO and OH lines and
$K$-band spectra CN lines.  In both these ranges numerous, strong atomic
lines are present.  These lines were used to determine abundances of carbon,
nitrogen, and oxygen and elements around the iron peak: Sc, Ti, Fe, Ni.  The
$K_{\rm r}$-band spectra are dominated by strong CO features that enable
measurement of the $^{12}$C$/^{13}$C isotopic ratio.  The spectra were
extracted and wavelength calibrated using standard reduction techniques
\citep{Joy1992}.  The wavelength scales of all spectra were heliocentric
corrected.  In all cases, telluric lines were either not present in the
interval observed or were removed by reference to a hot standard star.  The
Gaussian instrumental profile is in all cases about 6\,km\,s$^{-1}$ FWHM
(full width at half-maximum) corresponding to instrumental profiles of 0.31,
0.44, and 0.47\AA\ in the case of the $H$-, $K$-, and $K_{\rm r}$-band
spectra, respectively.  The journal of our spectroscopic observations is
given in the Table\,\ref{T1}.

\begin{table}
\centering   
\caption{Quadrature sums of the projected rotational velocities and
microturbulence $(V_{\rmn{rot}}^2 \sin^2{i} + \xi^2_{\rmn{t}})^{0.5}$ from
$K$-band \mbox{Ti\,{\sc i}}, \mbox{Fe\,{\sc i}}, and \mbox{Sc\,{\sc i}}
lines.$\,^a$}\label{T3}
\begin{tabular}{@{}lcccc@{}} \hline
                      & AE\,Ara          & BX\,Mon         & KX\,TrA         & CL\,Sco        \\
\hline
 Apr\,2003            & 10.06 $\pm$ 0.32 & 8.67 $\pm$ 0.47 & 8.48 $\pm$ 0.59 & 7.84 $\pm$ 0.60\\
 Aug\,2003            & --               & --              & --              & 8.02 $\pm$ 0.54\\
 Apr\,2004            & 10.58 $\pm$ 0.48 & --              & 8.94 $\pm$ 0.71 & 8.42 $\pm$ 0.50\\
 all together$\,^{b}$ & 10.30 $\pm$ 0.28 & 8.67 $\pm$ 0.47 & 8.71 $\pm$ 0.44 & 8.09 $\pm$ 0.29\\
\hline
\end{tabular}
\begin{list}{}{}
\item[$^{a}$]\,Units $\rmn{km}\,\rmn{s}^{-1}$
\item[$^{b}$]\,Used for synthetic spectra calculations
\end{list}
 \end{table}

\section{Methods}\label{secmet}

Abundance analyses were performed by fitting synthetic spectra to the
observed spectra using the same methods adopted in Paper\,I for the analyses
of RW\,Hya and SY\,Mus.  The technique is very similar to that used by
\cite{Sch2006} in determining the CH\,Cyg abundances.  Standard LTE analysis
and {\small{MARCS}} model atmospheres by \cite{Gus2008} were used for the
spectral synthesis.  The code {\small{WIDMO}} developed by \citet{Sch2006},
was used to calculate synthetic spectra over the entire observed spectral
region.  To perform the $\chi^2$ minimization a special overlay was
developed on the {\small{WIDMO}} code with use of the simplex algorithm
\citep{Bra1998}.  This procedure enables an improvement of the computation
efficiency by a factor of ten.  The atomic data were taken from the VALD
database \citep{Kup1999} in the case of $K$- and $K_{\rm r}$-band regions
and from the list given by \cite{MeBa1999} for the $H$-band region.  For the
molecular data we used the lists \cite{Goo1994} for CO and \citet{Kur1999}
for $^{12}$CN and OH.  The complete list of the lines selected for our
abundance analysis with excitation potentials (EP) and $gf$-values is
shown in Tables\,\ref{TLA} and \ref{TLM} in the online Appendix\,B.

The input effective temperatures $T_{\rm{eff}}$ were estimated
(Table\,\ref{T2}) from the known spectral types \citep{Mue1999} adopting the
calibrations by \cite{Ric1999} and \cite{VBe1999}.  The infrared intrinsic
colours derived from the 2MASS \citep{Phi2007} magnitudes and colour
excesses (\citealt{Sch2011}, \citealt{Sch1998}) combined with the
\cite{Kuc2005} $T_{\rm{eff}}$--$\log{g}$--colour relation for late-type
giants resulted in surface gravities and effective temperatures that are in
good agreement with those from the spectral types.  The adopted model
atmospheres had effective temperatures $T_{\rm{eff}}= 3300$\,K for AE\,Ara
and KX\,TrA and 3400\,K for CL\,Sco and BX\,Mon.  $\log{g} = 0$ with the
exception of CL\,Sco where $\log{g}$ was set to $0.5$.  It is difficult to
estimate uncertainty in the adopted $\log{g}$, however, it should not be
larger than $\sim 0.5$ which is the resolution of the {\small{MARCS}} model
atmosphere grid used in our calculation.  In the case of BX\,Mon, an
additional constraint on the $\log{g}$ can be obtained using the red giant
mass, $M_{\rm g}=3.7\pm1.9\, \rm M\sun$, and radius, $R_{\rm g}=160\pm50\,
\rm R\sun$ derived by \citet{Dum1998}.  The resulting
$\log{g}=0.6^{+0.5}_{-0.6}$.  Using the significantly improved orbital
solution \citep{Bra2009}, the red giant mass is $M_{\rm g}\sim1.5\, \rm
M\sun$, and the resulting $\log{g}\sim0.2$, in good agreement with the
value(s) adopted in our study.  The macroturbulence velocity
$\zeta_{\rm{t}}$ was set at 3\,km$/$s, a value typical for the cool red
giants \citep[e.g.][]{Fek2003}.

To obtain radial and rotational velocities, we used a cross-correlation
technique similar to that adopted by \cite{Car2011} but using synthetic
spectra as the templates.  The method is described in detail in Paper\,I and
the obtained values are presented in Table\,\ref{T1}.  The rotational
velocities were additionally estimated (Table\,\ref{T3}) via direct
measurement of the FWHM of the six relatively strong unblended atomic lines
(\mbox{Ti\,{\sc i}}, \mbox{Fe\,{\sc i}}, \mbox{Sc\,{\sc i}}) present in the
$K$-band region.  The same lines were used by \cite{Fek2003} to measure the
rotational velocities in roughly a dozen symbiotic systems.  We used the
radial velocities obtained by cross-correlation (Table\,\ref{T1}) and
rotational velocities obtained from atomic lines in $K$-band spectra
(Table\,\ref{T3}) as fixed parameters in our solutions.

A detailed description of the methods used to estimate the input parameters
and to derive the abundance solution was presented in Paper\,I.  A brief
outline follows.  Values of the abundance parameters (C, N, O, Sc, Ti, Fe,
Ni) were initially set to the solar composition \citep{Asp2009}.  Abundances
of the oxygen, carbon, nitrogen, and iron peak elements were adjusted by
fitting by eye, alternately from the OH, CO, CN, and atomic lines, over
several iterations.  Next, the initial grid of the $n+1$, $n$ dimensional
sets of free parameters, the so-called simplex needed for the simplex
algorithm, was prepared.  Nine different simplexes were used with different
microturbulent velocity $\xi_{\rm{t}}$ values sampled in the range
1.2--2.6\,km$/$s to obtain best fits to $H$- and $K$-band spectra.  For
three objects (AE\,Ara, KX\,TrA, and CL\,Sco) for which we do not have the
$K_{\rm r}$-band spectrum the $^{12}$C$/^{13}$C isotopic ratio was set to 8,
a value close to the average for our objects with known isotopic ratios. 
For BX\,Mon after we found the sets of parameters that give the best fit to
the $H$- and $K$-band spectra, we applied these abundances to the $K_{\rm
r}$-band spectrum as a fixed parameter and searched for $^{12}$C$/^{13}$C
isotopic ratio.  After obtaining the optimal fit, we made a reconciliation
of $^{12}$C and $^{12}$C$/^{13}$C in three iterations.

\begin{figure}
  \includegraphics[width=84mm]{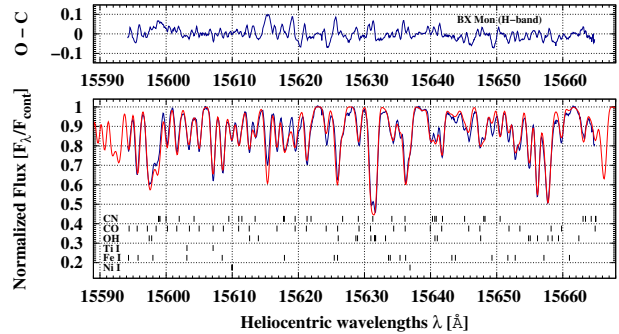}
  \caption{The spectrum of BX\,Mon observed in 2003\,February (blue line) in
the $H$\,band and a synthetic spectrum (red line) calculated using the final
abundances and $^{12}$C/$^{13}$C isotopic ratio (Table\,\ref{T4}).}
  \label{F1}
\end{figure}

\begin{figure}
  \includegraphics[width=84mm]{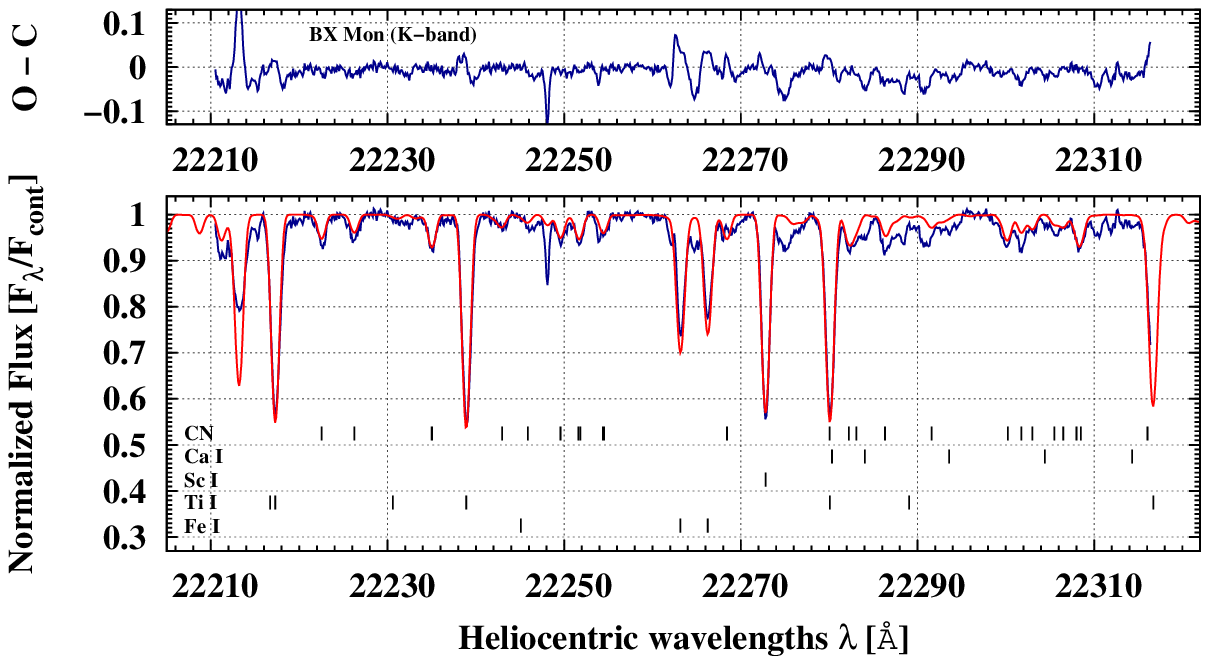} 
  \caption{The spectrum of BX\,Mon observed in 2003\,April (blue line) in
the $K$\,band and a synthetic spectrum (red line) calculated using the final
abundances and $^{12}$C/$^{13}$C isotopic ratio (Table\,\ref{T4}).}
  \label{F2}
\end{figure}

\begin{figure}
  \includegraphics[width=84mm]{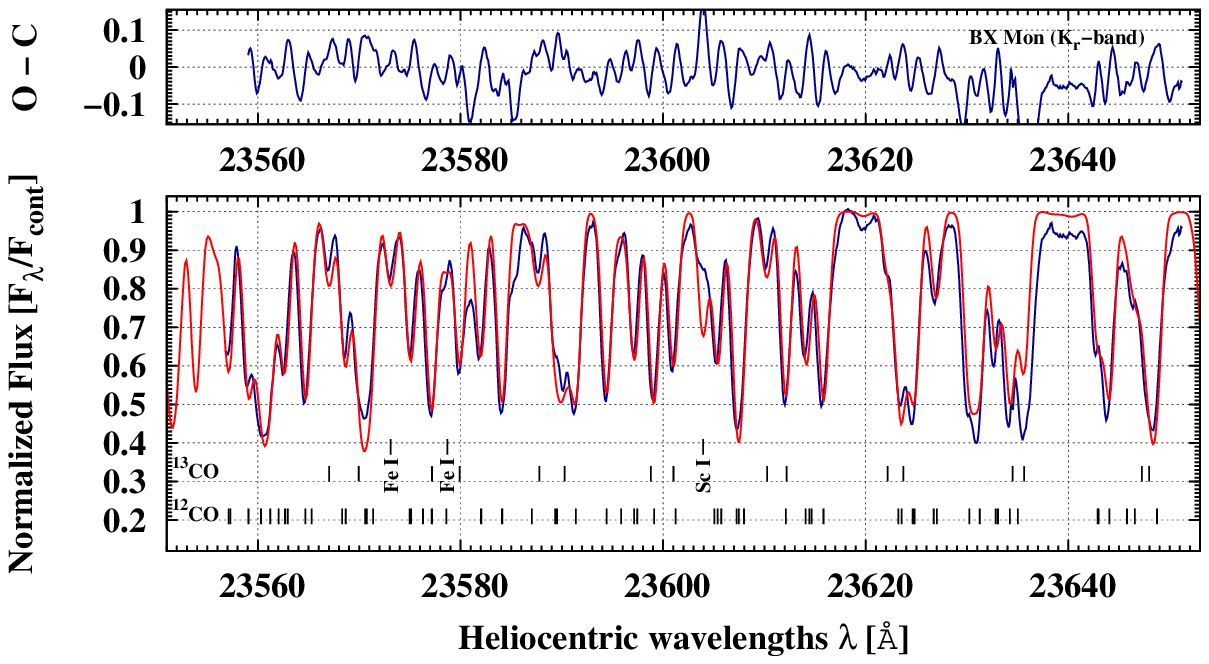} 
  \caption{The spectrum of BX\,Mon observed in 2006\,April (blue line) in
the $K_{\rm r}$\,band and a synthetic spectrum (red line) calculated using the
final abundances and $^{12}$C/$^{13}$C isotopic ratio (Table\,\ref{T4}).}
  \label{F3}
\end{figure}

\begin{table*}
 \centering
  \caption{Calculated abundances and relative abundances,$\,^a$ velocity
parameters,$\,^b$ and uncertainties$\,^c$ for AE\,Ara, BX\,Mon, KX\,TrA, and
CL\,Sco.}
  \label{T4}
  \begin{tabular}{@{}lcccccccc@{}}
  \hline
                        &\multicolumn{2}{c}{AE\,Ara}     &\multicolumn{2}{c}{BX\,Mon}      &\multicolumn{2}{c}{KX\,TrA}      &\multicolumn{2}{c}{CL\,Sco}     \\
  $X$                   & $\log{\epsilon(X)}$  & [$X$]   & $\log{\epsilon(X)}$ & [$X$]     & $\log{\epsilon(X)}$ & [$X$]     & $\log{\epsilon(X)}$  & [$X$]   \\
  \hline
  $^{12}$C              & 8.10$\pm$0.02& -0.33$\pm$0.07  & 7.79$\pm$0.02  & -0.64$\pm$0.07 & 8.03$\pm$0.02  & -0.40$\pm$0.07 & 8.02$\pm$0.04& -0.41$\pm$0.09  \\
  N                     & 8.15$\pm$0.06& +0.32$\pm$0.11  & 7.89$\pm$0.04  & +0.06$\pm$0.09 & 8.04$\pm$0.06  & +0.21$\pm$0.11 & 8.14$\pm$0.13& +0.31$\pm$0.18  \\
  O                     & 8.64$\pm$0.04& -0.05$\pm$0.08  & 8.37$\pm$0.03  & -0.32$\pm$0.07 & 8.66$\pm$0.05  & -0.03$\pm$0.10 & 8.61$\pm$0.06& -0.08$\pm$0.11  \\
  Sc                    & 4.53$\pm$0.24& +1.38$\pm$0.28  & 3.82$\pm$0.26  & +0.67$\pm$0.30 & 4.02$\pm$0.18  & +0.87$\pm$0.22 & 3.47$\pm$0.25& +0.32$\pm$0.29  \\
  Ti                    & 5.40$\pm$0.12& +0.45$\pm$0.17  & 4.96$\pm$0.15  & +0.01$\pm$0.20 & 5.08$\pm$0.17  & +0.13$\pm$0.22 & 4.93$\pm$0.22& -0.02$\pm$0.27  \\
  Fe                    & 7.41$\pm$0.06& -0.09$\pm$0.10  & 7.16$\pm$0.06  & -0.34$\pm$0.10 & 7.17$\pm$0.03  & -0.33$\pm$0.07 & 7.21$\pm$0.09& -0.29$\pm$0.13  \\
  Ni                    & 6.28$\pm$0.18& +0.06$\pm$0.22  & 6.18$\pm$0.13  & -0.04$\pm$0.17 & 6.21$\pm$0.09  & -0.01$\pm$0.13 & 6.23$\pm$0.10& +0.01$\pm$0.14  \\
  $^{12}$C/$^{13}$C     & --           & --              & 8$\pm$1        & --             & --             & --             & --           & --              \\
  \hline
$\xi_{\rmn{t}}$         &\multicolumn{2}{c}{1.7$\pm$0.2} &\multicolumn{2}{c}{1.8$\pm$0.2}  &\multicolumn{2}{c}{1.9$\pm$0.2}  &\multicolumn{2}{c}{1.9$\pm$0.3} \\
$V_{\rmn{rot}} \sin{i}$ &\multicolumn{2}{c}{10.2$\pm$0.3}&\multicolumn{2}{c}{8.5$\pm$0.5}  &\multicolumn{2}{c}{8.5$\pm$0.5}  &\multicolumn{2}{c}{7.9$\pm$0.3} \\
  \hline
\end{tabular}
\begin{list}{}{}
\item[$^a$] Relative to the Sun [$X$] abundances estimated in relation to the solar composition of \citet{Asp2009}
\item[$^b$] Units $\rmn{km}\,\rmn{s}^{-1}$ 
\item[$^c$] 3$\sigma$
\end{list}
\end{table*}

\begin{table}
 \centering
  \caption{Sensitivity of abundances to uncertainties in the stellar parameters}
\label{T5}
  \begin{tabular}{@{}lcccc@{}}

  \hline
  $\Delta X$ & $\Delta T_{\rmn{eff}} = +100$\,K & $\Delta \log{g} = +0.5$ & $\Delta  \xi_{\rmn{t}} = +0.5$ & $\Delta [Fe/H] = +0.25$ \\
  \hline
  C              & +0.02 & +0.22 & -0.08 & +0.05 \\
  N              & +0.04 & +0.02 & -0.11 & +0.07 \\
  O              & +0.13 & +0.08 & -0.12 & +0.12 \\
  Sc             & +0.12 & +0.20 & -0.69 & -0.01 \\
  Ti             & +0.07 & +0.16 & -0.52 & +0.02 \\
  Fe             & -0.05 & +0.16 & -0.17 & -0.02 \\
  Ni             & -0.07 & +0.20 & -0.22 & -0.02 \\
  \hline
\end{tabular}
\end{table}

\section{Results}\label{secres}

Table\,\ref{T4} summarizes the final abundances and formal uncertainties
derived from CNO molecules and atomic lines (\mbox{Sc\,{\sc i}},
\mbox{Ti\,{\sc i}}, \mbox{Fe\,{\sc i}}, \mbox{Ni\,{\sc i}}) on the scale of
$\log{\epsilon}(X) = \log{N(X) N(H)^{-1}} + 12.0$, the isotopic ratio
$^{12}$C/$^{13}$C (only in the case of BX\,Mon), microturbulences,
$\xi_{\rm{t}}$, and projected rotational velocities, $V_{\rm{rot}} \sin{i}$. 
The abundance of scandium is based on only one strong \mbox{Sc\,{\sc i}}
line at $\lambda \sim 22272.8$\AA\ and may be less reliable than other
abundances presumably due to the broadening of the infrared scandium lines
by hyperfine structure that has not been included in the analysis (see
Paper\,I).  The synthetic fits to the observed spectra of BX\,Mon are shown
in Figures\,\ref{F1}-\ref{F3}, where the molecular (OH, CO, CN) and atomic
(\mbox{Sc\,{\sc i}}, \mbox{Ti\,{\sc i}}, \mbox{Fe\,{\sc i}}, \mbox{Ni\,{\sc
i}}) lines used in the solution of the chemical composition are identified
by ticks.  Synthetic fits to the observed spectra of AE\,Ara, KX\,TrA, and
CL\,Sco in $H$- and $K$-band regions are shown in
Figures\,\ref{FA1}-\ref{FA6} in the online Appendix\,A.

We also made fits with {\small{MARCS}} atmosphere models varying the
effective temperatures by $\pm$100\,K, $\log{g}$ by $\pm$0.5, and the
microturbulence ($\xi_{\rm t}$) by $\pm$0.5 to estimate the sensitivity of
abundances to uncertainties in stellar parameters.  In the case of AE\,Ara
and CL\,Sco additional fits were also made using models with [Fe/H]
different by +0.25.  The changes in the abundance for each element as a
function of each model parameter are listed in Table\,\ref{T5}. 
Uncertainties in the stellar parameters have a stronger impact on the
uncertainty of the derived chemical composition than the uncertainties from
fitting the synthetic spectrum.

In our solution, the microturbulence velocity was treated as free parameter. 
In all cases, we obtained microturbulence values close to 2.  Abundances of
some elements strongly depend on the microturbulence value.  \citet{Smi2002}
showed that for red giants microturbulence is correlated with bolometric
magnitudes.  We repeated the analysis for BX\,Mon, AE\,Ara, KX\,TrA, and
CL\,Sco adopting the microturbulence values, 2.5, 2.5, 2.5, and 2.3,
respectively, corresponding to their bolometric magnitudes
(Table\,\ref{T2}).  The resulting abundances and their sensitivity on the
changes in the stellar parameters are presented in
Tables\,\ref{T6}\,and\,\ref{T7}, respectively.  These new abundances are
shifted towards somewhat lower (by $\sim$ 0.2\,dex) metallicities.  In
particular, for all objects we obtained sub-solar metallicities, [Fe$/$H]
$\sim -0.3$ to $-0.55$.  For BX\,Mon, we also calculated abundances for
$\log{g} = +0.5$.

\begin{table*}
 \centering
  \caption{Calculated abundances and relative abundances,$\,^a$ velocity
parameters,$\,^b$ and uncertainties$\,^c$ for AE\,Ara, BX\,Mon, KX\,TrA, and
CL\,Sco.  The case for fixed microturbulences.}
  \label{T6}
  \begin{tabular}{@{}lcccccccccc@{}}
  \hline
                        &\multicolumn{2}{c}{AE\,Ara}     &\multicolumn{2}{c}{BX\,Mon}      &\multicolumn{2}{c}{BX\,Mon$^{d}$}&\multicolumn{2}{c}{KX\,TrA}      &\multicolumn{2}{c}{CL\,Sco}     \\
  $X$                   & $\log{\epsilon(X)}$  & [$X$]   & $\log{\epsilon(X)}$ & [$X$]     & $\log{\epsilon(X)}$ & [$X$]     & $\log{\epsilon(X)}$ & [$X$]     & $\log{\epsilon(X)}$  & [$X$]   \\
  \hline
  $^{12}$C              & 7.93$\pm$0.03& -0.50$\pm$0.08  & 7.66$\pm$0.03  & -0.77$\pm$0.08 & 7.87$\pm$0.03  & -0.56$\pm$0.08 & 7.89$\pm$0.03  & -0.54$\pm$0.07 & 7.92$\pm$0.04& -0.51$\pm$0.09  \\
  N                     & 7.89$\pm$0.09& +0.06$\pm$0.14  & 7.65$\pm$0.06  & -0.18$\pm$0.11 & 7.67$\pm$0.05  & -0.16$\pm$0.10 & 7.82$\pm$0.09  & -0.01$\pm$0.14 & 7.96$\pm$0.10& +0.13$\pm$0.15  \\
  O                     & 8.35$\pm$0.03& -0.34$\pm$0.07  & 8.14$\pm$0.01  & -0.55$\pm$0.06 & 8.23$\pm$0.02  & -0.46$\pm$0.06 & 8.39$\pm$0.05  & -0.30$\pm$0.10 & 8.39$\pm$0.03& -0.30$\pm$0.08  \\
  Sc                    & 3.45$\pm$0.22& +0.30$\pm$0.25  & 3.15$\pm$0.22  & +0.00$\pm$0.26 & 3.29$\pm$0.18  & +0.14$\pm$0.22 & 3.33$\pm$0.12  & +0.18$\pm$0.16 & 3.16$\pm$0.23& +0.01$\pm$0.27  \\
  Ti                    & 4.55$\pm$0.08& -0.40$\pm$0.13  & 4.33$\pm$0.06  & -0.62$\pm$0.11 & 4.48$\pm$0.07  & -0.47$\pm$0.12 & 4.46$\pm$0.12  & -0.49$\pm$0.17 & 4.53$\pm$0.12& -0.42$\pm$0.17  \\
  Fe                    & 7.16$\pm$0.06& -0.34$\pm$0.10  & 6.95$\pm$0.04  & -0.55$\pm$0.08 & 7.10$\pm$0.04  & -0.40$\pm$0.08 & 6.96$\pm$0.06  & -0.54$\pm$0.10 & 7.07$\pm$0.08& -0.43$\pm$0.12  \\
  Ni                    & 5.95$\pm$0.09& -0.27$\pm$0.13  & 5.93$\pm$0.10  & -0.29$\pm$0.14 & 6.09$\pm$0.08  & -0.13$\pm$0.12 & 5.96$\pm$0.08  & -0.26$\pm$0.12 & 6.11$\pm$0.10& -0.11$\pm$0.14  \\
  $^{12}$C/$^{13}$C     & --           & --              & 12$\pm$2       &  --            & 11$\pm$1       & --             & --             & --             & --           & --              \\
  \hline
$\xi_{\rmn{t}}$         &\multicolumn{2}{c}{2.5}         &\multicolumn{2}{c}{2.5}          &\multicolumn{2}{c}{2.5}          &\multicolumn{2}{c}{2.5}          &\multicolumn{2}{c}{2.3        } \\
$V_{\rmn{rot}} \sin{i}$ &\multicolumn{2}{c}{10.0$\pm$0.3}&\multicolumn{2}{c}{8.3$\pm$0.5}  &\multicolumn{2}{c}{8.3$\pm$0.5}  &\multicolumn{2}{c}{8.3$\pm$0.5}  &\multicolumn{2}{c}{7.8$\pm$0.3} \\
  \hline
\end{tabular}
\begin{list}{}{}
\item[$^a$] Relative to the Sun [$X$] abundances estimated in relation to the solar composition of \citet{Asp2009}
\item[$^b$] Units $\rmn{km}\,\rmn{s}^{-1}$
\item[$^c$] 3$\sigma$
\item[$^d$] $\log{g} = 0.5$
\end{list}
\end{table*}

\section{Discussion}\label{secdiscussion}

Here, we present the first ever analysis of the photospheric chemical
abundances (CNO and elements around the iron peak: Sc, Ti, Fe, and Ni) for
four classical S-type symbiotic systems: AE\,Ara, BX\,Mon, KX\,TrA, and
CL\,Sco.  Our analysis reveals an approximately solar metallicity for
AE\,Ara and slightly sub-solar metallicities ([Fe$/$H] $\sim -0.3$) for
BX\,Mon, KX\,TrA, and CL\,Sco.  The CNO abundances are similar to typical
values derived for single Galactic M giants.  In particular, they all show
carbon depletion and nitrogen enhancement \citep{SmLa1990}.  The ratio of
$^{12}$C$/^{13}$C $\sim$ 8 obtained for BX\,Mon is very similar to the
values of $^{12}$C$/^{13}$C $\sim$ 6 and 10 derived for RW\,Hya and SY\,Mus
(Paper\,I).  The CNO values and isotopic ratios indicate that the red giants
in these systems have experienced the first dredge-up.

Relative abundances of C/N/O for these systems were previously derived based
on nebular emission lines in ultraviolet spectra by \cite{Nus1988} and
\cite{Per1995}.  \cite{Per1995} also estimated Fe/O from optical emission
lines.  A comparison of these estimates with our present results and those
from Paper\,I is shown in Table\,\ref{T8}.  The abundances from emission
lines should be most reliable when based on spectra taken during superior
conjunction of the cool giant ($\phi \sim 0.5$) when the hot component and
the nebula are visible in front of the giant.  This was the case for
KX\,TrA, AE\,Ara, and BX\,Mon and perhaps SY\,Mus.  The C/O ratios obtained
using emission line technique are in fairly good agreement with our
photospheric values especially for KX\,TrA and AE\,Ara.  For the other
objects the differences are bigger.  In the case of BX\,Mon, the difference
is likely due to poor quality of the {\it IUE} spectrum used by
\citet{Nus1988}.  The {\it IUE} spectra of CL\,Sco and RW\,Hya were taken
relatively close to the inferior conjunction of the red giant when the
emission lines are affected by eclipse and Rayleigh scattering effects. 
\citet{Nus1988} also noted that their method may systematically
underestimate the C and O abundances relative to the N abundance and this
effect in the worst cases may reach 30 per cent.

\begin{table}
 \centering
  \caption{Sensitivity of abundances to uncertainties in the stellar
parameters.  The case for fixed microturbulences.}
\label{T7}
  \begin{tabular}{@{}lccc@{}}

  \hline
  $\Delta X$ & $\Delta T_{\rmn{eff}} = +100$\,K & $\Delta \log{g} = +0.5$ & $\Delta  \xi_{\rmn{t}} = +0.5$ \\
  \hline
  C              & +0.02 & +0.22 & -0.04 \\
  N              & +0.02 & +0.03 & -0.12 \\
  O              & +0.10 & +0.10 & -0.09 \\
  Sc             & +0.10 & +0.16 & -0.38 \\
  Ti             & +0.06 & +0.17 & -0.39 \\
  Fe             & -0.05 & +0.17 & -0.19 \\
  Ni             & -0.05 & +0.18 & -0.17 \\
  \hline
\end{tabular}
\end{table}

\begin{figure}
  \resizebox{\hsize}{!}{\includegraphics[]{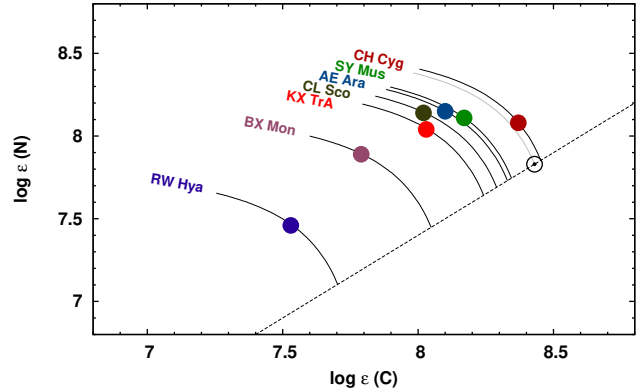}}
  \caption{Nitrogen versus carbon for the symbiotic giants.  The dashed line
represents scaled solar abundances, [$^{12}$C/Fe]=0 and [$^{14}$N/Fe]=0,
whereas the solid curves delineate constant $^{12}$C+$^{14}$N.}
  \label{F10}
\end{figure}

The elemental abundances of the symbiotic giants summarized in
Table\,\ref{T9} can be used to address evolutionary status and to associate
symbiotic systems with stellar populations of the Milky Way.  In particular
the C and N abundances are very good monitors of dredge-up on the RGB
provided that only the CN cycle has operated significantly in the dredged
material.  In such a case, the total number of C+N nuclei should be
conserved since $^{12}$C will be converted to $^{14}$N.  \citet{CuSm2006}
showed that the Galactic bulge giants behave in good agreement with the
simple CN mixing picture.  Paper\,I demonstrated the same behaviour of the
red giants in RW\,Hya, SY\,Mus, and CH\,Cyg.  Fig.\,\ref{F10} show the
$^{14}$N versus $^{12}$C abundances for the symbiotic sample so far
analysed.  All symbiotic giants fall into the $^{14}$N-enhanced zone
providing a very strong indication that they have experienced the first
dredge-up.  This is also confirmed by the low $^{12}$C/$^{13}$C isotopic
ratios for those stars with measured values.

Thus, we can assume, as did \citet{CuSm2006} for their sample of the bulge
giants, that the O and Fe and other elemental abundances roughly represent
the original values with which the symbiotic stars were born.  These
abundances can then be used to identify the parent Galactic population.  The
ratio of $\alpha$-elements (e.g.  O and Ti in our study) to Fe is of
particular importance because the $\alpha$-element originate mostly from
massive stars and SNe\,II, and they are produced over relatively short
time-scales whereas Fe is most effectively produced in SNe\,Ia over much
longer time-scales.  Thus, differences in contamination by SNe II and SNe Ia
lead to significantly different trends for particular populations, and their
clear separation in the [O$/$Fe] versus [Fe$/$H] plane (e.g. 
\citealt{CuSm2006}, \citealt{Ben2006}).

\begin{table}
\centering   
\caption{Comparison of our photospheric abundances with those derived using emission line technique.}
\label{T8}
\begin{tabular}{@{}lllllll@{}} 
\hline
Object   & C/N  & O/N  & C/O  & Fe/O  & Phase$^{a}$  & Ref.  \\
\hline
KX\,TrA  & 1.59 & 4.42 & 0.36 & 0.046 & 0.52                & $[3]$ \\
         & 2.28 & 6.91 & 0.33 & --    & 0.27                & $[3]$ \\
         & 0.98 & 4.17 & 0.23 & 0.032 &                     & $[1]$ \\
AE\,Ara  & 1.2  & 3.7  & 0.32 & --    & 0.43                & $[4]$ \\
         & 0.89 & 3.09 & 0.29 & 0.059 &                     & $[1]$ \\
BX\,Mon  & 0.29 & 2.0  & 0.15 & --    & 0.49                & $[4]$ \\
         & 0.79 & 3.02 & 0.26 & 0.062 &                     & $[1]$ \\
CL\,Sco  & 0.77 & 5.7  & 0.14 & --    & 0.87                & $[4]$ \\
         & 0.76 & 2.95 & 0.26 & 0.040 &                     & $[1]$ \\
SY\,Mus  & 0.97 & 1.9  & 0.51 & --    & 0.36                & $[4]$ \\
         & 1.15 & 3.55 & 0.32 & 0.058 &                     & $[2]$ \\
RW\,Hya  & 0.42 & 1.7  & 0.25 & --    & 0.15                & $[4]$ \\
         & 1.17 & 5.13 & 0.23 & 0.037 &                     & $[2]$ \\
\hline
\end{tabular} 
\begin{list}{}{}
\item[{\bf References:}] $^{[1]}$\,this work, $^{[2]}$\,Paper\,I, $^{[3]}$\,\cite{Per1995}, $^{[4]}$\,\cite{Nus1988}.
\item[$^{a}$] Orbital phase the same as in Table\,\ref{T1}, and table 2 in Paper\,I.

\end{list}
\end{table}

\begin{table}
 \centering
  \caption{Absolute and relative abundances adopted for the comparison with
the Galactic stellar populations.  Two cases for the microturbulence being
free and fixed parameter are shown at the top and at the middle,
respectively.  Abundances of RW\,Hya and SY\,Mus from Paper I and CH\,Cyg and V2116\,Oph from 
the literature are shown for comparison at the bottom.}
  \label{T9}
  \begin{tabular}{@{}lccccl@{}}
  \hline
   Object           & $A$($^{12}$C) & $A$($^{14}$N) & $[$O$/$Fe$]$ & $[$Fe$/$H$]$ & $\xi_{\rmn{t}}$ \\
  \hline
AE\,Ara             & 8.10 & 8.15 & +0.04 & -0.09 & 1.7\\
                    & 7.93 & 7.89 & +0.00 & -0.34 & 2.5\\
BX\,Mon             & 7.79 & 7.89 & +0.02 & -0.34 & 1.8\\
                    & 7.66 & 7.65 & +0.00 & -0.55 & 2.5\\
                    & 7.87 & 7.67 & -0.06 & -0.40 & 2.5$^{a}$\\
KX\,TrA             & 8.03 & 8.04 & +0.30 & -0.33 & 1.9\\
                    & 7.89 & 7.82 & +0.24 & -0.54 & 2.5\\
CL\,Sco             & 8.02 & 8.14 & +0.21 & -0.29 & 1.9\\
                    & 7.92 & 7.96 & +0.13 & -0.43 & 2.3\\
RW\,Hya$^{[1]}$     & 7.53 & 7.46 & +0.24 & -0.76 & 1.8\\
SY\,Mus$^{[1]}$     & 8.17 & 8.11 & +0.05 & -0.08 & 2.0\\
CH\,Cyg$^{[2]}$     & 8.37 & 8.08 & +0.07 & +0.00 & 2.2\\
V2116\,Oph$^{[3]}$  & 8.03 & 8.97 & -0.22 & -0.05 & 2.4\\
  \hline
\end{tabular}
\begin{list}{}{}
\item[{\bf References:}] $^{[1]}$\,Paper\,I, $^{[2]}$\,\cite{Sch2006}, $[3]$\,\cite{Hin2006}.
\item[$^{a}$] $\log{g} = 0.5$
\end{list}
\end{table}

\begin{figure}
  \resizebox{\hsize}{!}{\includegraphics[]{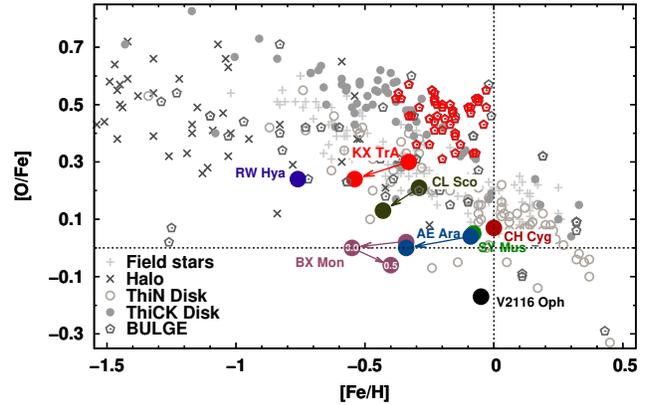}}
  \caption{Oxygen relative to iron for various stellar populations with
positions of our targets denoted with large coloured circles.  The results
of additional calculations with fixed microturbulence values are designated
by arrows (for BX\,Mon two values of $\log{g} = 0$ and $+0.5$ were used).}
  \label{F11}
\end{figure}

Figure\,\ref{F11} shows [O$/$Fe] versus  [Fe$/$H] of the symbiotic giants 
along with the values for various stellar populations taken from a number of
studies (\citealt{Edv1993}, \citealt{Pro2000}, \citealt{Mel2001},
\citealt{Smi2001b}, \citealt{MeBa2002}, \citealt{Ful2003},
\citealt{Red2003}, \citealt{Ric2005}, \citealt{Ben2005}, \citealt{CuSm2006},
\citealt{Alv2010}, \citealt{Ryd2010}, \citealt{Ric2012}, \citealt{Smi2013})
and scaled to the solar composition of \cite{Asp2009}.  We have
distinguished four populations: thin and thick discs, halo, and bulge.  Part
of the bulge population containing objects from Baade's Windows and from
two other nearby inner bulge fields F175 and F265 (\citealt{Ric2005};
\citealt{Ric2012}) is highlighted with red.  These objects are more
chemically homogeneous than the rest of the bulge sample as indicated by
their grouping in a small area of the diagram.  The rest of the sample
contains objects spread in the bulge (\citealt{CuSm2006,Alv2010,Ryd2010})
and is more representative for the whole bulge population.  The thin and
thick disc samples contain only those objects with membership confirmed by
their kinematic characteristics.  Finally, the 'field star' group represents
the objects for which there is no kinematic information about their
population membership.  It may contain stars from all Galactic populations
but it seems to be dominated by the thin disc members.

The position in the [O/Fe]--[Fe/H] plane of the objects studied here can be
related to membership in Galactic populations.  Most appear to belong to the
Galactic disc or bulge while the position of RW\,Hya supports its membership
in the extended thick-disc/halo population.  Additional independent methods,
e.g.  simultaneous kinematic studies on the Toomre diagram
(\citealt{Fel2003, Gal2014}), are necessary to sort out this question.  Such
studies will be performed in the near future on a more statistically
significant sample.

In the process of fitting the synthetic spectra we measured radial and
rotational velocities for the programme stars (Tables\,\ref{T1} and
\ref{T3}).  Radial velocities obtained for three objects, AE\,Ara, BX\,Mon,
and CL\,Sco, using cross-correlation techniques are consistent with recent
spectroscopic orbits published for these stars.  Radial velocities obtained
for AE\,Ara are in agreement with results obtained with the same spectra by
\citet{Fek2010} with discrepancies no larger than 2 km s$^{-1}$.  Residuals
calculated by \citet{Fek2010} from the synthetic orbit are within the range
$\sim$1.5--2 km s$^{-1}$.  Similarly the radial velocities obtained for
CL\,Sco agree with values obtained by the \citet{Fek2007} with the same
spectra with differences in the range $\sim$0.2--2.3 km s$^{-1}$.  Residuals
from the synthetic orbit calculated by \citet{Fek2007} have values between
$\sim$1-- and 3 km s$^{-1}$.  Radial velocities obtained for BX\,Mon are in
accord with synthetic radial velocities predicted from the orbit of
\citet{Fek2000} with an accuracy generally better than $\sim$2 km s$^{-1}$. 
In the case of KX\,TrA, however, we could not achieve agreement with
published orbits.  The range of the velocities measured by us, -122 to -126
km s$^{-1}$, is covered by the $\gamma$ velocity of -123.7 km s$^{-1}$ and
$K$ of 6.8 km s$^{-1}$ found by \citet{Fer2003} but computed versus observed
velocities are not consistent with an orbital period of 1350\,d.  The orbit
of \citet{HaHo2000} from spectropolarimetric observation is similar to that
of \citet{Fer2003}.  The \citet{Mar2008} orbit is similar but the 1916\,d
period is significantly longer.  Further study and possibly recalculation
with another period is needed.

Giants in symbiotic stars are characterized with large rotational velocities
and in all the systems studied by us so far $V_{\rm{rot}}\sin{i}$ makes the
largest contribution to the physical line broadening.  The rotational
velocities obtained with the cross-correlation method generally have smaller
values than obtained with FWHM method (Tables\,\ref{T1} and \ref{T3}).  The
rotational velocities obtained from strongly blended spectra with strong
molecular lines (like $H$-band spectra) appear to be significantly
underestimated, and the use of rotational velocities as a free parameters
does not lead to significant differences in obtained values of the
rotational velocities and abundances (Paper\,I).  Therefore, our analysis
used velocities obtained with FWHM method that gives more self-consistent
results in accord with those obtained from the same spectra with
cross-correlation method.

\section{Conclusions}

We have performed a detailed analysis of the photospheric abundances
of CNO and elements around the iron peak (Sc, Ti, Fe, and Ni) for
the red giant components of the S-type symbiotic binaries: AE\,Ara,
BX\,Mon, KX\,TrA, and CL\,Sco.  Our analysis revealed a near-solar
metallicity for AE\,Ara, and slightly sub-solar metallicities
([Fe$/$H]$\sim-0.3$) for BX\,Mon, KX\,TrA, and CL\,Sco.  However, the
metallicities are lower by $\sim 0.2$\,dex when the microturbulence
values were estimated from bolometric magnitudes instead of keeping
the microturbulence as a free parameter.  The enrichment in $^{14}$N
isotope obtained for all these objects indicates that the giants
have experienced the first dredge-up which is also confirmed by
the very low $^{12}$C/$^{13}$C isotopic ratio $\sim$8 obtained for
BX\,Mon.

\section*{Acknowledgements}

This study has been supported in part by the Polish NCN grant no.
DEC-2011/01/B/ST9/06145.  CG has been also financed by the NCN postdoc
programme FUGA via grant DEC-2013/08/S/ST9/00581.  The observations were
obtained at the Gemini Observatory, which is operated by the Association of
Universities for Research in Astronomy, Inc., under a cooperative agreement
with the NSF on behalf of the Gemini partnership: the National Science
Foundation (United States), the National Research Council (Canada), CONICYT
(Chile), the Australian Research Council (Australia), Minist\'{e}rio da
Ci\^{e}ncia, Tecnologia e Inova\c{c}\~{a}o (Brazil) and Ministerio de
Ciencia, Tecnolog\'{i}a e Innovaci\'{o}n Productiva (Argentina).  
CG is thankful
to Miros{\l}aw Schmidt for his valuable and useful comments.

                                     
\appendix

\newpage

\section[]{Observed and synthetic spectra of AE\,Ara, KX\,TrA, and CL\,Sco}

\newpage

\begin{figure}
  \includegraphics[width=84mm]{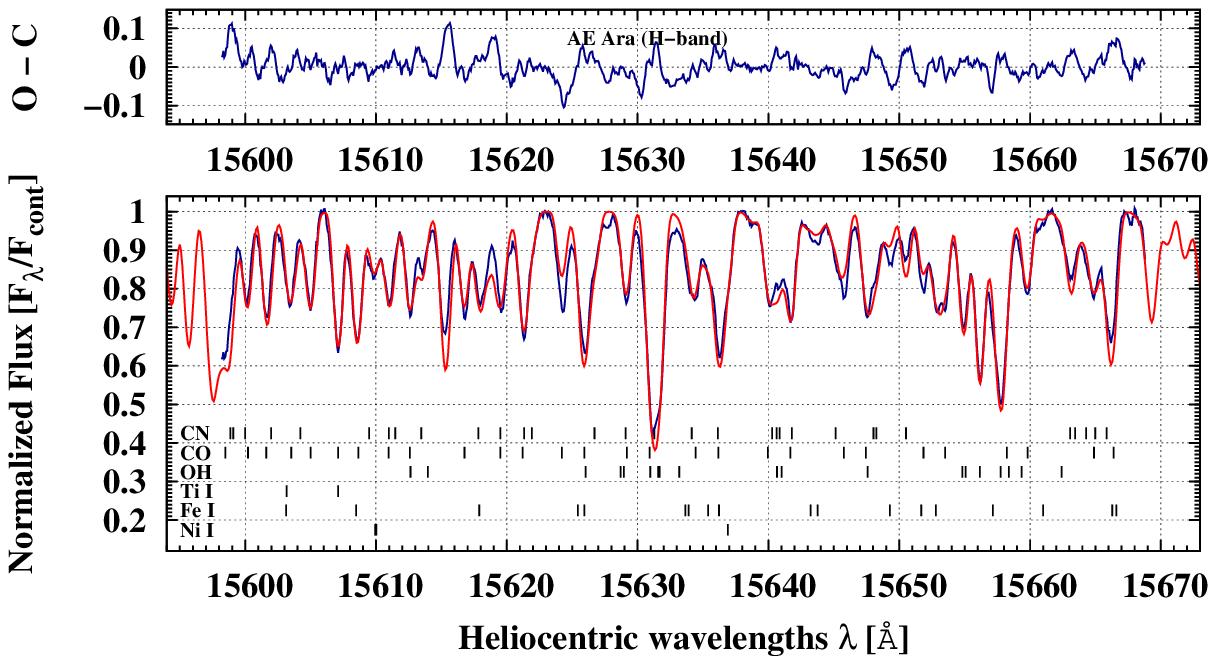}
  \caption{The spectrum of AE\,Ara observed in 2003\,February (blue line) in
the $H$ band and a synthetic spectrum (red line) calculated using the final
abundances (Table\,\ref{T4}).}
  \label{FA1}
\end{figure}

\begin{figure}
  \includegraphics[width=84mm]{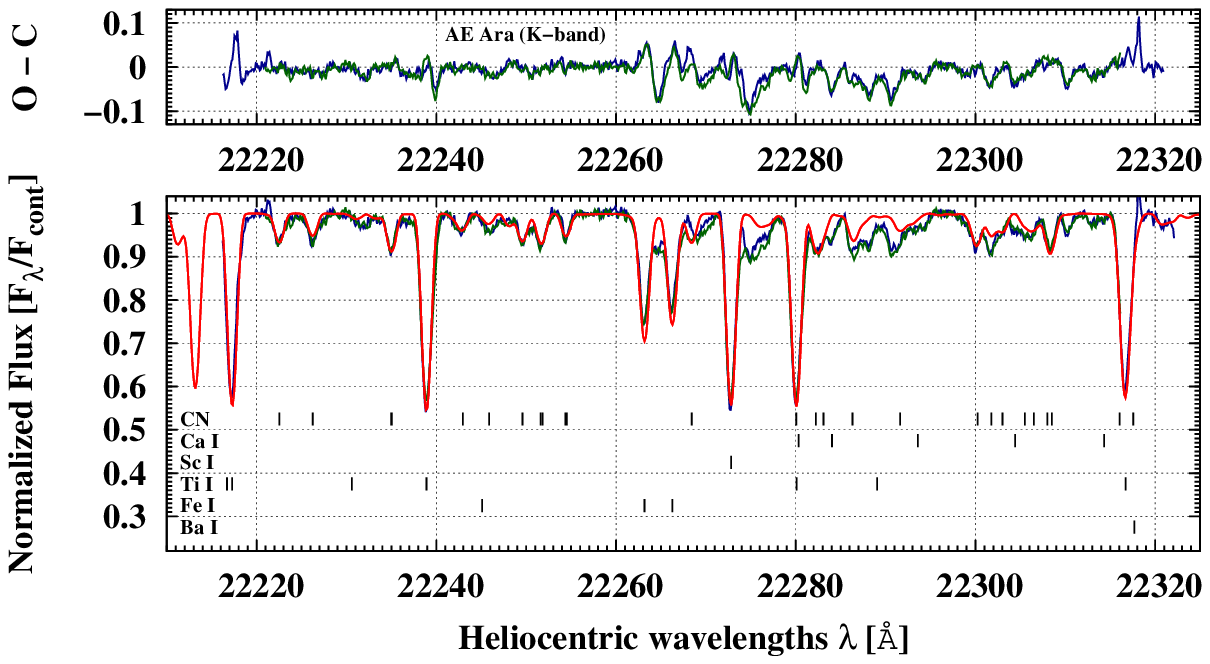} 
  \caption{Spectra of AE\,Ara observed in 2003\,April (blue line) and
April\,2004 (green line) in the $K$ band and a synthetic spectrum (red line)
calculated using the final abundances (Table\,\ref{T4}).}
  \label{FA2}
\end{figure}

\begin{figure}
  \includegraphics[width=84mm]{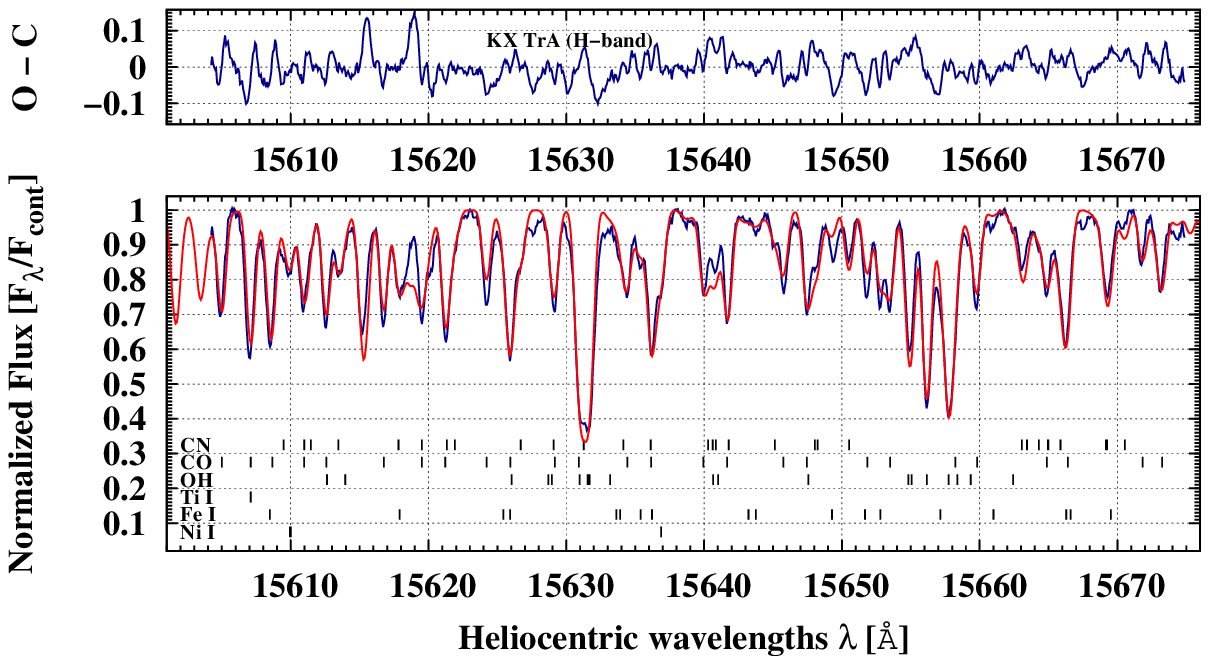}
  \caption{The spectrum of KX\,TrA observed in 2003\,February (blue line) in
the $H$ band and a synthetic spectrum (red line) calculated using the final
abundances (Table\,\ref{T4}).}
  \label{FA3}
\end{figure}

\begin{figure}
  \includegraphics[width=84mm]{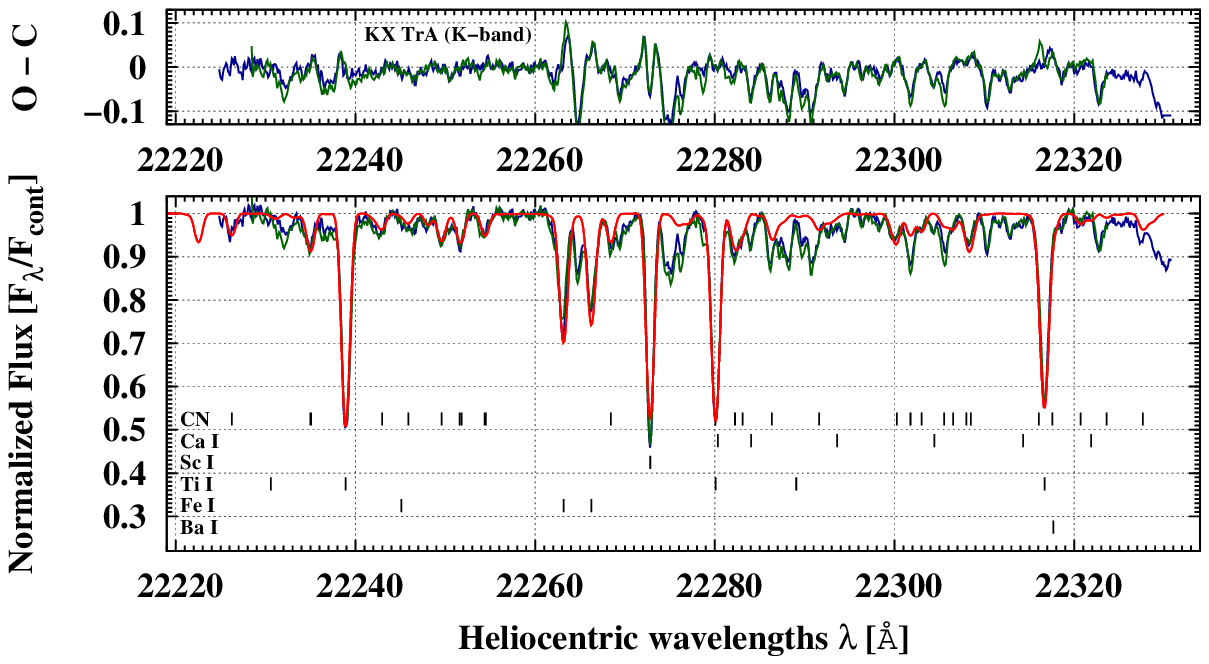} 
  \caption{Spectra of KX\,TrA observed in 2003\,April (blue line) and
April\,2004 (green line) in the $K$ band and a synthetic spectrum (red line)
calculated using the final abundances (Table\,\ref{T4}).}
  \label{FA4}
\end{figure}

\begin{figure}
  \includegraphics[width=84mm]{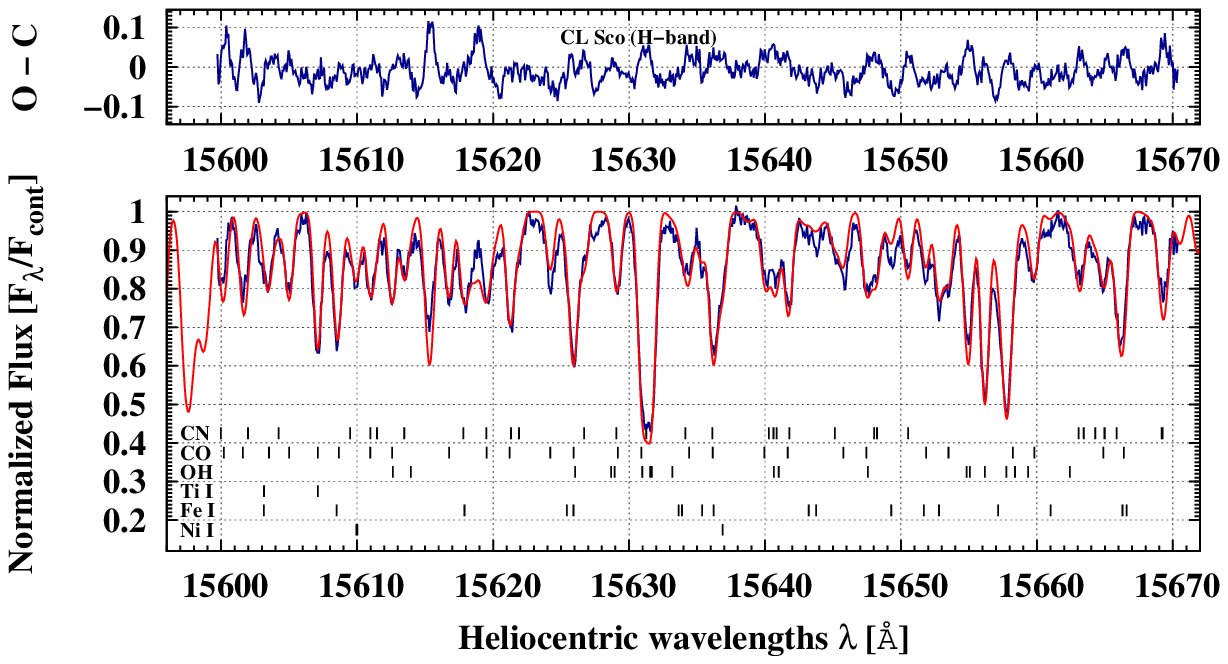}
  \caption{The spectrum of CL\,Sco observed in 2003\,February (blue line) in
the $H$ band and a synthetic spectrum (red line) calculated using the final
abundances (Table\,\ref{T4}).}
  \label{FA5}
\end{figure}

\begin{figure}
  \includegraphics[width=84mm]{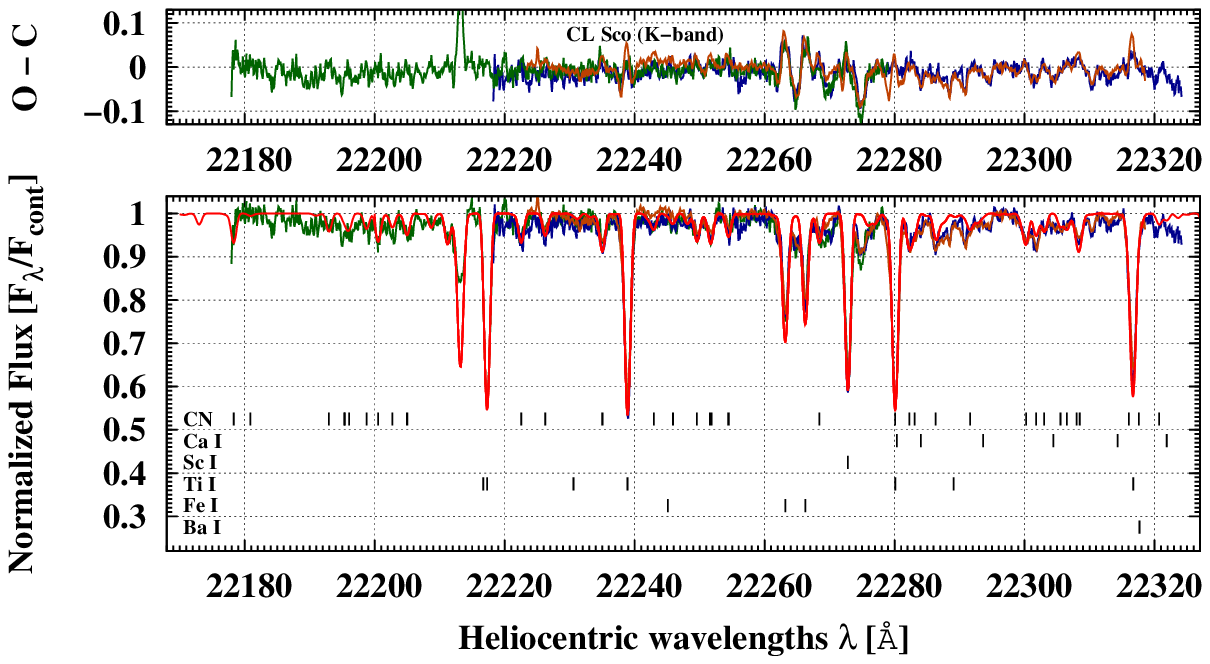} 
  \caption{Spectra of CL\,Sco observed in 2003\,April (blue line),
2003\,August (green line), and April\,2004 (dark-orange line) in the $K$ band
and a synthetic spectrum (red line) calculated using the final abundances
(Table\,\ref{T4}).}
  \label{FA6}
\end{figure}


\section[]{List of atomic and molecular lines}

\begin{table}
 \centering
  \caption{List of atomic lines selected for calculations
together with $gf$-values and excitation potentials.}\label{TLA}
  \begin{tabular}{@{}lccc@{}}
  \hline
Wavelength (air) & $EP$     & $\log{gf}$ & Ref. \\
$(\AA)$    & $($eV$)$ &            &      \\
  \hline
\mbox{Sc\,{\sc i}}&          &            & \\
$^{a}$22202.640  & 1.428    & -3.436     & \citet{Kup1999}\\
$^{a}$22207.128  & 1.428    & -1.836     & \citet{Kup1999}\\
22266.729  & 1.428    & -1.327     & \citet{Kup1999}\\
$^{a}$ 23603.945 & 1.428    & -2.180     & \citet{Kup1999}\\
\mbox{Ti\,{\sc i}}&          &            & \\
15598.890  & 4.690    & -0.030     & \citet{MeBa1999}\\
15602.840  & 2.270    & -1.810     & \citet{MeBa1999}\\
22210.636  & 4.213    & -1.444     & \citet{Kup1999}\\
22211.228  & 1.734    & -1.770     & \citet{Kup1999}\\
22224.530  & 4.933    & -0.263     & \citet{Kup1999}\\
22232.838  & 1.739    & -1.658     & \citet{Kup1999}\\
22274.012  & 1.749    & -1.756     & \citet{Kup1999}\\
22282.973  & 4.690    & -0.586     & \citet{Kup1999}\\
22310.617  & 1.734    & -2.124     & \citet{Kup1999}\\
\mbox{Fe\,{\sc i}}&          &            & \\
15590.050  & 6.240    & -0.550     & \citet{MeBa1999}\\
15591.490  & 6.240    &  0.360     & \citet{MeBa1999}\\
15593.740  & 5.030    & -1.980     & \citet{MeBa1999}\\
15598.870  & 6.240    & -0.920     & \citet{MeBa1999}\\
15604.220  & 6.240    &  0.280     & \citet{MeBa1999}\\
15613.630  & 6.350    & -0.290     & \citet{MeBa1999}\\
15621.160  & 6.200    & -0.990     & \citet{MeBa1999}\\
15621.650  & 5.540    & +0.170     & \citet{MeBa1999}\\
15629.370  & 5.950    & -1.670     & \citet{MeBa1999}\\
15629.630  & 4.560    & -3.130     & \citet{MeBa1999}\\
15631.110  & 3.640    & -3.980     & \citet{MeBa1999}\\
15631.950  & 5.350    & -0.150     & \citet{MeBa1999}\\
15638.950  & 5.810    & -1.810     & \citet{MeBa1999}\\
15639.480  & 6.410    & -0.870     & \citet{MeBa1999}\\
15645.010  & 6.310    & -0.540     & \citet{MeBa1999}\\
15647.410  & 6.330    & -1.090     & \citet{MeBa1999}\\
15648.520  & 5.430    & -0.800     & \citet{MeBa1999}\\
15652.870  & 6.250    & -0.190     & \citet{MeBa1999}\\
15656.640  & 5.870    & -1.900     & \citet{MeBa1999}\\
15662.010  & 5.830    & +0.000     & \citet{MeBa1999}\\
15662.320  & 6.330    & -1.020     & \citet{MeBa1999}\\
15665.240  & 5.980    & -0.600     & \citet{MeBa1999}\\
22239.040  & 5.385    & -3.121     & \citet{Kup1999}\\
22257.097  & 5.064    & -0.723     & \citet{Kup1999}\\
22260.185  & 5.086    & -0.952     & \citet{Kup1999}\\
$^{a}$23566.671  & 6.144    &  0.306     & \citet{Kup1999}\\
$^{a}$23572.267  & 5.874    & -0.886     & \citet{Kup1999}\\
\mbox{Ni\,{\sc i}}&          &            & \\
15605.6800 & 5.300    & -0.590     & \citet{MeBa1999}\\
15605.7500 & 5.300    & -1.010     & \citet{MeBa1999}\\
15632.6200 & 5.310    & -0.130     & \citet{MeBa1999}\\
  \hline                             
\end{tabular}
\begin{list}{}{}
\item[$^{a}$] Not used for the chemical composition determination.
\end{list}                       
\end{table}                          

\begin{table}
 \centering  
  \caption{List of molecular lines selected for calculations
together with $gf$-values and excitation potentials.}\label{TLM}
\begin{tabular}{@{}lccc@{}}
  \hline
Wavelength (air) & $EP$     & $\log{gf}$ & Ref. \\
$(\AA)$    & $($eV$)$ &            &      \\
  \hline
$^{12}$C$^{14}$N     &          &            & \\
15594.607  & 0.9162   & -1.645     & \citet{Kur1999}\\%
15594.610  & 1.1079   & -1.240     & \citet{Kur1999}\\%
15594.816  & 1.1770   & -1.097     & \citet{Kur1999}\\%
15595.729  & 0.9218   & -2.106     & \citet{Kur1999}\\%
15597.719  & 1.3900   & -1.443     & \citet{Kur1999}\\%
15599.961  & 1.2391   & -1.807     & \citet{Kur1999}\\%
15605.217  & 1.2601   & -1.810     & \citet{Kur1999}\\%
15606.718  & 0.9095   & -2.116     & \citet{Kur1999}\\%
15607.198  & 1.5181   & -1.318     & \citet{Kur1999}\\%
15609.198  & 0.9361   & -1.479     & \citet{Kur1999}\\%
15613.554  & 1.0521   & -1.645     & \citet{Kur1999}\\%
15613.557  & 1.1770   & -1.090     & \citet{Kur1999}\\%
15615.246  & 1.0359   & -1.645     & \citet{Kur1999}\\%
15617.057  & 1.1080   & -1.231     & \citet{Kur1999}\\%
15617.646  & 1.3900   & -1.435     & \citet{Kur1999}\\%
15622.431  & 0.9162   & -1.637     & \citet{Kur1999}\\%
15624.806  & 1.5182   & -1.311     & \citet{Kur1999}\\%
15626.993  & 0.9345   & -2.088     & \citet{Kur1999}\\%
15629.867  & 1.2602   & -1.798     & \citet{Kur1999}\\%
15631.866  & 0.9362   & -1.473     & \citet{Kur1999}\\%
15636.015  & 0.9218   & -2.097     & \citet{Kur1999}\\%
15636.361  & 1.2816   & -1.801     & \citet{Kur1999}\\%
15636.584  & 1.1314   & -1.231     & \citet{Kur1999}\\%
15637.521  & 0.9357   & -1.633     & \citet{Kur1999}\\%
15640.865  & 1.4182   & -1.436     & \citet{Kur1999}\\%
15643.764  & 1.0688   & -1.632     & \citet{Kur1999}\\%
15643.965  & 1.0521   & -1.632     & \citet{Kur1999}\\%
15646.246  & 1.2059   & -1.090     & \citet{Kur1999}\\%
15658.785  & 1.1314   & -1.223     & \citet{Kur1999}\\%
15659.161  & 0.9476   & -2.070     & \citet{Kur1999}\\%
15660.014  & 1.5516   & -1.313     & \citet{Kur1999}\\%
15660.688  & 1.4182   & -1.428     & \citet{Kur1999}\\%
15660.705  & 1.2817   & -1.789     & \citet{Kur1999}\\%
15661.584  & 0.9612   & -1.471     & \citet{Kur1999}\\%
15664.879  & 1.2059   & -1.083     & \citet{Kur1999}\\
15664.966  & 0.9358   & -1.626     & \citet{Kur1999}\\
15666.270  & 0.9345   & -2.079     & \citet{Kur1999}\\
22172.254  & 1.2026   & -1.762     & \citet{Kur1999}\\
22174.832  & 1.0475   & -3.212     & \citet{Kur1999}\\
22186.906  & 1.0849   & -2.206     & \citet{Kur1999}\\
22189.293  & 1.0975   & -2.868     & \citet{Kur1999}\\
22189.324  & 1.7232   & -2.022     & \citet{Kur1999}\\
22190.012  & 1.0975   & -2.273     & \citet{Kur1999}\\
22192.684  & 1.1177   & -2.235     & \citet{Kur1999}\\
22194.477  & 1.2309   & -1.747     & \citet{Kur1999}\\%
22196.680  & 1.4573   & -1.875     & \citet{Kur1999}\\%
22198.914  & 1.4373   & -1.872     & \citet{Kur1999}\\%
22198.955  & 1.6007   & -1.825     & \citet{Kur1999}\\%

22216.477  & 1.2165   & -1.747     & \citet{Kur1999}\\%
22220.176  & 1.1036   & -2.055     & \citet{Kur1999}\\%
22228.924  & 1.1079   & -2.204     & \citet{Kur1999}\\%
22228.994  & 1.3036   & -1.633     & \citet{Kur1999}\\%
22236.898  & 1.1074   & -2.249     & \citet{Kur1999}\\%
22239.814  & 1.6007   & -1.816     & \citet{Kur1999}\\%
22243.514  & 1.2456   & -1.732     & \citet{Kur1999}\\%
22245.525  & 1.3622   & -1.803     & \citet{Kur1999}\\
22245.752  & 1.1285   & -2.213     & \citet{Kur1999}\\
22248.297  & 1.4573   & -1.862     & \citet{Kur1999}\\%
22248.438  & 1.4777   & -1.866     & \citet{Kur1999}\\%
22262.354  & 1.2309   & -1.732     & \citet{Kur1999}\\%
22273.969  & 1.6282   & -1.818     & \citet{Kur1999}\\%
  \hline                             
\end{tabular}                        
\end{table}

\begin{table}                           
 \centering                             
  \contcaption{}                        
\begin{tabular}{@{}lccc@{}}             
  \hline
Wavelength (air) & $EP$     & $\log{gf}$ & Ref. \\
$(\AA)$    & $($eV$)$ &            &      \\
  \hline
22276.160  & 1.3036   & -1.624     & \citet{Kur1999}\\%
22277.008  & 1.1080   & -2.197     & \citet{Kur1999}\\%
22280.246  & 1.1036   & -2.047     & \citet{Kur1999}\\%
22285.527  & 1.1177   & -2.225     & \citet{Kur1999}\\%
22294.160  & 1.2609   & -1.718     & \citet{Kur1999}\\%
22295.695  & 1.1216   & -2.042     & \citet{Kur1999}\\%
22296.930  & 1.3900   & -1.803     & \citet{Kur1999}\\%
22299.426  & 1.4778   & -1.853     & \citet{Kur1999}\\%
22300.430  & 1.1397   & -2.192     & \citet{Kur1999}\\%
22301.910  & 1.4986   & -1.856     & \citet{Kur1999}\\%
22302.410  & 1.3260   & -1.624     & \citet{Kur1999}\\%
22309.961  & 1.2457   & -1.718     & \citet{Kur1999}\\%
22311.473  & 1.2942   & -2.377     & \citet{Kur1999}\\%
22314.611  & 1.6282   & -1.809     & \citet{Kur1999}\\
22317.520  & 1.7561   & -2.009     & \citet{Kur1999}\\
22321.525  & 1.1314   & -2.195     & \citet{Kur1999}\\
$^{12}$C$^{16}$O     &          &            & \\
15590.144  & 0.4690   & -7.3583    & \citet{Goo1994}\\
15591.363  & 0.1427   & -7.7310    & \citet{Goo1994}\\
15592.908  & 0.4901   & -7.3428    & \citet{Goo1994}\\
15594.223  & 0.1313   & -7.7545    & \citet{Goo1994}\\
15595.946  & 0.5118   & -7.3273    & \citet{Goo1994}\\
15597.348  & 0.1204   & -7.7786    & \citet{Goo1994}\\
15599.257  & 0.5339   & -7.3121    & \citet{Goo1994}\\
15600.737  & 0.1100   & -7.8035    & \citet{Goo1994}\\
15602.842  & 0.5564   & -7.2971    & \citet{Goo1994}\\
15604.392  & 0.1000   & -7.8294    & \citet{Goo1994}\\
15606.702  & 0.5794   & -7.2822    & \citet{Goo1994}\\
15608.312  & 0.0905   & -7.8564    & \citet{Goo1994}\\
15612.497  & 0.0814   & -7.8841    & \citet{Goo1994}\\
15615.250  & 0.6268   & -7.2530    & \citet{Goo1994}\\
15616.948  & 0.0729   & -7.9133    & \citet{Goo1994}\\
15619.940  & 0.6512   & -7.2385    & \citet{Goo1994}\\
15621.663  & 0.0648   & -7.9439    & \citet{Goo1994}\\
15624.908  & 0.6760   & -7.2242    & \citet{Goo1994}\\
15626.644  & 0.0572   & -7.9755    & \citet{Goo1994}\\
15630.155  & 0.7012   & -7.2101    & \citet{Goo1994}\\
15631.891  & 0.0500   & -8.0092    & \citet{Goo1994}\\
15635.682  & 0.7269   & -7.1961    & \citet{Goo1994}\\
15637.404  & 0.0434   & -8.0446    & \citet{Goo1994}\\
15641.490  & 0.7531   & -7.1822    & \citet{Goo1994}\\
15643.182  & 0.0372   & -8.0824    & \citet{Goo1994}\\
15647.580  & 0.7797   & -7.1685    & \citet{Goo1994}\\
15649.227  & 0.0314   & -8.1226    & \citet{Goo1994}\\
15653.953  & 0.8068   & -7.1548    & \citet{Goo1994}\\
15655.537  & 0.0262   & -8.1659    & \citet{Goo1994}\\
15660.610  & 0.8343   & -7.1412    & \citet{Goo1994}\\
15662.115  & 0.0214   & -8.2128    & \citet{Goo1994}\\
15667.552  & 0.8622   & -7.1277    & \citet{Goo1994}\\
15668.960  & 0.0172   & -8.2640    & \citet{Goo1994}\\
23550.868  & 1.2764   & -4.9838    & \citet{Goo1994}\\
23550.681  & 1.4307   & -4.2927    & \citet{Goo1994}\\
23552.653  & 0.8557   & -4.5830    & \citet{Goo1994}\\
23553.886  & 0.2917   & -5.4983    & \citet{Goo1994}\\
23554.786  & 0.0048   & -6.4587    & \citet{Goo1994}\\
23556.234  & 1.4594   & -4.2830    & \citet{Goo1994}\\
23558.269  & 0.8385   & -4.5974    & \citet{Goo1994}\\
23562.231  & 1.4886   & -4.2734    & \citet{Goo1994}\\
23564.307  & 0.8218   & -4.6121    & \citet{Goo1994}\\
23564.951  & 2.1495   & -4.3503    & \citet{Goo1994}\\
23568.674  & 1.5182   & -4.2638    & \citet{Goo1994}\\
23569.863  & 2.5598   & -4.7140    & \citet{Goo1994}\\
23570.723  & 0.2870   & -5.5423    & \citet{Goo1994}\\
  \hline                             
\end{tabular}                        
\end{table}

\begin{table}
 \centering  
  \contcaption{}
\begin{tabular}{@{}lccc@{}}
  \hline
Wavelength (air) & $EP$     & $\log{gf}$ & Ref. \\
$(\AA)$    & $($eV$)$ &            &      \\
  \hline
23570.766  & 0.8056   & -4.6271    & \citet{Goo1994}\\
23575.563  & 1.5482   & -4.2545    & \citet{Goo1994}\\
23577.647  & 0.7898   & -4.6424    & \citet{Goo1994}\\
23577.693  & 0.0071   & -6.3642    & \citet{Goo1994}\\
23582.901  & 1.5786   & -4.2451    & \citet{Goo1994}\\
23582.995  & 2.1902   & -4.3426    & \citet{Goo1994}\\
23584.949  & 0.7744   & -4.6582    & \citet{Goo1994}\\
23587.982  & 0.2827   & -5.5907    & \citet{Goo1994}\\
23590.689  & 1.6095   & -4.2359    & \citet{Goo1994}\\
23592.672  & 0.7596   & -4.6743    & \citet{Goo1994}\\
23594.794  & 2.6067   & -4.7071    & \citet{Goo1994}\\
23598.928  & 1.6408   & -4.2268    & \citet{Goo1994}\\
23600.817  & 0.7451   & -4.6908    & \citet{Goo1994}\\
23601.032  & 0.0100   & -6.2876    & \citet{Goo1994}\\
23601.535  & 2.2314   & -4.3350    & \citet{Goo1994}\\
23605.665  & 0.2789   & -5.6445    & \citet{Goo1994}\\
23607.621  & 1.6725   & -4.2177    & \citet{Goo1994}\\
23609.383  & 0.7312   & -4.7077    & \citet{Goo1994}\\
23616.77   & 1.7047   & -4.2088    & \citet{Goo1994}\\
23618.369  & 0.7177   & -4.7251    & \citet{Goo1994}\\
23620.259  & 2.6541   & -4.7001    & \citet{Goo1994}\\
23620.571  & 2.2729   & -4.3273    & \citet{Goo1994}\\
23623.771  & 0.2756   & -5.7051    & \citet{Goo1994}\\
23624.802  & 0.0133   & -6.2232    & \citet{Goo1994}\\
23626.375  & 1.7373   & -4.2000    & \citet{Goo1994}\\
23627.778  & 0.7046   & -4.7430    & \citet{Goo1994}\\
23636.44   & 1.7703   & -4.1912    & \citet{Goo1994}\\
23637.608  & 0.6921   & -4.7617    & \citet{Goo1994}\\
23640.108  & 2.3148   & -4.3198    & \citet{Goo1994}\\
23642.301  & 0.2728   & -5.7747    & \citet{Goo1994}\\
$^{12}$C$^{17}$O     &          &            & \\
23555.604  & 0.5678   & -4.9176    & \citet{Goo1994}\\
23558.862  & 0.0255   & -5.9974    & \citet{Goo1994}\\
23561.873  & 0.5513   & -4.9326    & \citet{Goo1994}\\
23568.546  & 0.5353   & -4.9473    & \citet{Goo1994}\\
23575.626  & 0.5198   & -4.9630    & \citet{Goo1994}\\
23583.110  & 0.5047   & -4.9788    & \citet{Goo1994}\\
23591.000  & 0.4901   & -4.9948    & \citet{Goo1994}\\
23599.294  & 0.4759   & -5.0114    & \citet{Goo1994}\\
23607.995  & 0.4621   & -5.0284    & \citet{Goo1994}\\
23617.100  & 0.4489   & -5.0459    & \citet{Goo1994}\\
23626.613  & 0.4360   & -5.0639    & \citet{Goo1994}\\
23636.530  & 0.4237   & -5.0825    & \citet{Goo1994}\\
$^{12}$C$^{18}$O     &          &            & \\
23556.518  & 0.2538   & -5.4594    & \citet{Goo1994}\\
23564.155  & 0.2389   & -5.4752    & \citet{Goo1994}\\
23572.184  & 0.2245   & -5.4913    & \citet{Goo1994}\\
23580.602  & 0.2105   & -5.5079    & \citet{Goo1994}\\
23589.412  & 0.1970   & -5.5251    & \citet{Goo1994}\\
23598.614  & 0.1839   & -5.5426    & \citet{Goo1994}\\
23608.206  & 0.1712   & -5.5607    & \citet{Goo1994}\\
23618.19   & 0.1590   & -5.5792    & \citet{Goo1994}\\
23628.565  & 0.1473   & -5.5984    & \citet{Goo1994}\\
23639.333  & 0.1360   & -5.6182    & \citet{Goo1994}\\
$^{13}$C$^{16}$O     &          &            & \\
23560.596  & 1.3099   & -4.9751    & \citet{Goo1994}\\
23563.521  & 0.1719   & -5.5580    & \citet{Goo1994}\\
23570.753  & 1.3437   & -4.9666    & \citet{Goo1994}\\
23573.485  & 0.1596   & -5.5766    & \citet{Goo1994}\\
23581.341  & 1.3779   & -4.9582    & \citet{Goo1994}\\
23583.842  & 0.1478   & -5.5957    & \citet{Goo1994}\\
23592.363  & 1.4126   & -4.9496    & \citet{Goo1994}\\
23594.59   & 0.1365   & -5.6156    & \citet{Goo1994}\\
  \hline                             
\end{tabular}                           
\end{table}

\begin{table}
 \centering  
  \contcaption{}
\begin{tabular}{@{}lccc@{}}
  \hline
Wavelength (air) & $EP$     & $\log{gf}$ & Ref. \\
$(\AA)$    & $($eV$)$ &            &      \\
  \hline
23603.818  & 1.4476   & -4.9412    & \citet{Goo1994}\\
23605.733  & 0.1256   & -5.6360    & \citet{Goo1994}\\
23615.711  & 1.4831   & -4.9329    & \citet{Goo1994}\\
23617.268  & 0.1151   & -5.6574    & \citet{Goo1994}\\
23628.043  & 1.5189   & -4.9248    & \citet{Goo1994}\\
23629.197  & 0.1051   & -5.6794    & \citet{Goo1994}\\
23640.814  & 1.5552   & -4.9165    & \citet{Goo1994}\\
23641.522  & 0.0956   & -5.7025    & \citet{Goo1994}\\
OH         &          &            & \\
15593.179  & 0.8740   & -5.358     & \citet{Kur1999}\\
15593.563  & 0.8740   & -5.358     & \citet{Kur1999}\\
15608.357  & 0.4942   & -7.209     & \citet{Kur1999}\\%
15609.683  & 0.4942   & -7.209     & \citet{Kur1999}\\%
15621.766  & 0.8374   & -6.734     & \citet{Kur1999}\\%
15624.434  & 0.8415   & -7.006     & \citet{Kur1999}\\%
15624.660  & 0.1336   & -8.233     & \citet{Kur1999}\\%
15626.704  & 0.5413   & -5.198     & \citet{Kur1999}\\%
15627.290  & 0.8871   & -5.435     & \citet{Kur1999}\\%
15627.293  & 0.8871   & -5.435     & \citet{Kur1999}\\%
15627.413  & 0.5413   & -5.198     & \citet{Kur1999}\\%
15628.902  & 0.1337   & -8.233     & \citet{Kur1999}\\%
15636.235  & 0.8876   & -7.202     & \citet{Kur1999}\\%
15636.596  & 0.8876   & -7.202     & \citet{Kur1999}\\%
15643.302  & 0.8420   & -7.007     & \citet{Kur1999}\\%
15650.557  & 0.8643   & -5.587     & \citet{Kur1999}\\%
15650.797  & 0.8643   & -5.587     & \citet{Kur1999}\\%
15651.896  & 0.5341   & -5.132     & \citet{Kur1999}\\%
15653.480  & 0.5343   & -5.132     & \citet{Kur1999}\\%
15654.116  & 0.8383   & -6.734     & \citet{Kur1999}\\%
15655.053  & 0.3041   & -7.713     & \citet{Kur1999}\\%
15658.127  & 0.3038   & -7.713     & \citet{Kur1999}\\%
  \hline                             
\end{tabular}                        
\end{table}                          
                                     
                                     
%
%


\bsp

\label{lastpage}

\end{document}